\documentclass[prb,showpacs,twocolumn,superscriptaddress,aps,a4paper]{revtex4-1}
\usepackage{dcolumn}
\usepackage{amsmath}
\usepackage{graphicx}
\usepackage{latexsym}
\usepackage{amsfonts}
\usepackage{amssymb}
\usepackage{color}
\DeclareGraphicsExtensions{.pdf,.gif,.jpg}

\def\a{\alpha}
\def\b{\beta}

\def\f{\phi}
\def\d{\delta}
\def\D{\Delta}

\def\e{\epsilon}

\def\vf{\varphi}

\def\inf{\infty}

\newcommand{\be}{\begin{equation}}
\newcommand{\ee}{\end{equation}}
\newcommand{\beq}{\begin{eqnarray}}
\newcommand{\eeq}{\end{eqnarray}}

\begin{document}




\title{On the thermalization of a Luttinger liquid after a sequence 
of sudden interaction quenches}



\author{E. Perfetto}
\affiliation{Dipartimento di Fisica, Universit\'a di Roma Tor
Vergata, Via della Ricerca Scientifica 1, I-00133 Rome, Italy}

\author{G. Stefanucci}
\affiliation{Dipartimento di Fisica, Universit\'a di Roma Tor
Vergata, Via della Ricerca Scientifica 1, I-00133 Rome, Italy}
\affiliation{INFN, Laboratori Nazionali di Frascati, Via E. Fermi 40, 00044 Frascati, Italy}
\affiliation{European Theoretical Spectroscopy Facility (ETSF)}





\begin{abstract}
    
We present a comprehensive analysis of the relaxation dynamics of a 
Luttinger liquid subject to a sequence of sudden interaction 
quenches.  We express  the 
critical exponent $\beta$ governing 
the decay of the steady-state propagator  as an
explicit  functional of the switching protocol. At long distances 
$\b$ depends only on the initial state while at short distances it 
is also history dependent. Continuous protocols of arbitrary 
complexity can be realized with  
infinitely long sequences. For quenches of finite duration 
we prove that there exist no protocol to bring the initial non-interacting system 
in the ground state of the Luttinger liquid. Nevertheless 
memory effects are  washed out at short-distances.
The adiabatic theorem is 
then investigated with ramp-switchings of increasing duration, and 
several analytic results for both the propagator and the excitation energy 
are derived.

\end{abstract}

\maketitle



 %
 %
 %
 %
 %
 %
 %
 %
 %
 %
 %
 %

\section{Introduction}

The on-going experimental activities on ultracold 
atoms are continuously challenging our 
understanding of many-body quantum systems.\cite{new}
Laser- or evaporative-cooled below the $\mu$K, atoms crystallize in 
artificial lattices\cite{rev} thus providing nearly ideal realizations of 
bosonic\cite{opt1,opt11,opt1a,opt1b} and fermionic\cite{opt22,opt2a,opt2b} model Hamiltonians.
The possibility of tuning the model parameters in {\em real time} 
brought the attention back to a fundamental issue in quantum 
statistical physics: does a non-interacting bulk system relax toward the 
correlated ground-state upon the switch-on of the interaction?

``Sudden quench'' is the nomenclature coined for the sudden change 
of a parameter like, e.g., the convexity of a parabolic trap or the 
interaction strength, in an equilibrium system.\cite{koll,man} During the last five 
years experimental and theoretical investigations on 
the relaxation properties of quenched ultracold atoms 
enlightened the intriguing phenomenon 
of the {\it thermalization breakdown}:\cite{thermb} either the system does not 
reach a steady state or, if it does, the steady state is not 
the ground state of the quenched Hamiltonian.
A sufficient criterion for the occurrence of a steady-state has been 
found, so far, only for integrable models\cite{bs.2008,stef,s.2007,c.2009} and it has been argued that
steady-state values are calculable by averaging over a generalized, 
initial-state dependent Gibbs 
ensemble.\cite{rigol1}

The thermalization breakdown 
poses  questions which are certainly conceptual in nature but may also be 
relevant  to the growing field of optimal control 
theory:\cite{wg.2007,bcr.2010,rc.2010} what is the steady-state 
dependence on the initial state? and on the switching protocol? how 
the adiabatic limit is recovered? In this paper we provide a 
comprehensive analysis of the behavior of a Luttinger Liquid (LL)  
subject to arbitrary {\em interaction quenches}. We extend the study of Cazalilla for 
a sudden quench\cite{cazalilla} to a sequence of $N$ sudden quenches using a recently 
proposed recursive method.\cite{perfetto} Continuous quenches of 
duration $T$ are then obtained in the limit $N\to\inf$ and allow
us to address the adiabatic limit by making $T$ larger and larger.  
We calculate the equal-time one-particle propagator 
$G^{[N]}(x,t)$ as well as the excitation energy.  
At long and short distances the steady-state $G^{[N]}(x,t)\sim x^{-\b}$ and in both 
cases we are able to write the critical exponent $\b$
as an explicit functional of the switching protocol. 
In the limit of continuous quenches ($N\to\inf$) the 
propagator at short distances thermalizes 
whereas at long distances does not. An analytic formula for ramp-like 
switchings of duration $T$ valid for all $x$ is derived and is shown that 
$G^{[\inf]}$ and the ground 
state propagator are  the same up to 
a critical distance that diverges for $T\to \inf$. 
The recovery of the adiabatic limit is further illustrated from  
energy balance considerations. The calculation of the excitation 
energy $\Delta E(T)$ is reduced to the solution of a simple differential equation 
that can be used to find the optimal switching protocol of duration 
$T$ that minimizes $\Delta E(T)$. We prove that $\Delta E(T)>0$ 
(strictly positive) for all switching protocols of finite duration 
and provide an analytic expression for ramp-like switchings.

The paper is organized as follows. In the next Section we introduce 
the model and the recursive procedure to calculate $G$ and $\D E$. 
Results for arbitrary sequences of sudden quenches are here 
illustrated.
The limit of continuous switching protocol is carried on in Section 
\ref{contprot} along with the derivation of several analytic 
formulas. A summary of the main findings is finally drawn in Section 
\ref{concsec}.


\section{Sequential quench in a Luttinger Liquid}

The sudden interaction quench in a LL has been addressed in a series of 
papers.\cite{cazalilla,perfetto1,cazalilla2,uhrig,zhou,zhou2,dhz.2010}
At the distance $x$ the propagator $G^{[1]}(x,t)$ exhibits the 
``light-cone'' effect, i.e., a crossover
between Fermi liquid behavior for times $t \ll x/2v$ ($v$ being the 
quasiparticle velocity) and nonthermal LL behavior in the long time limit. 
Sequential quenches yield an even richer phenomenology since additional time 
(and hence length) scales appear in the problem. We will show that 
different steady-state regimes emerge by probing the 
system at distances shorter or longer than the quenching time (in 
units of $v$) and that their nature depends on the switching protocol.

The LL Hamiltonian describes interacting spinless electrons confined in a 1D 
wire of length $\mathcal{L}$ and reads
\beq
H&=& \frac{1}{2}\sum_{\alpha=R,L} \int_{-\mathcal{L}/2}^{\mathcal{L}/2}dx 
[-2i \e_{\a} v_{F}\,
\psi^{\dagger}_{\alpha}(x)\partial_{x} \psi_{\alpha}(x) \nonumber \\
&+&
  g_{4}\rho^{2}_{\alpha}(x) + g_{2}  \rho_{\alpha}(x)
\rho_{\bar{\alpha}}(x)],
\label{model}
\eeq
where $\a$ denotes the chirality of the electrons with Fermi velocity 
$\e_{\a} v_{F}$ ($\e_{R/L}=\pm 1$)  and
density $\rho_{\alpha}=:\psi^{\dagger}_{\alpha} \psi_{\alpha}:$,  
``$:\;\;:$'' being the normal ordering.
The coupling constants $g_{2(4)}$ refer to forward scattering 
processes between electrons of opposite(identical) chirality.
We consider the system  noninteracting ($g_{2}=g_{4}=0$) and in the 
ground state before a series of 
$N$ interaction quenches at times $0 \equiv t_{0}<t_{1},\ldots < 
t_{N-1}\equiv T$ takes place. Let $g_{in}$ be the value of the couplings 
$g_{i=2,4}$ between $t_{n-1}$ and $t_{n}$ and $H_{n}$ the 
corresponding LL Hamiltonian. Each $H_{n}$ 
can be bosonized\cite{giamarchi} in terms 
of the scalar fields $\phi_{\a}$ defined from 
$\psi_{\a}(x)=\frac{\kappa_{\a}}{\sqrt{2\pi a} }
e^{i\sqrt{\pi}\phi_{\a}(x)}$, with $\kappa_{\a}$ the 
anticommuting Klein factors and $a$ a short-distance cutoff.
The  result is a simple quadratic form
\be
H_{n}=\frac{v_{n}}{2} \int_{-\infty}^{\infty}dx [K_{n}^{-1}
(\partial_{x}\phi(x))^{2}+K_{n} (\partial_{x}\theta(x))^{2}],
\label{scalar}
\ee
where $\phi=\phi_{R}+\phi_{L}$ and $\theta=\phi_{R}-\phi_{L}$ are 
conjugated fields, 
$v_{n}=\sqrt{(2\pi v_{F}+g_{4n})^{2}-g^{2}_{2n})}/2\pi$
is the renormalized velocity  and the parameter 
$K_{n}=\sqrt{(2\pi v_{F}+g_{4n}-g_{2n})/(2\pi v_{F}+g_{4n}+g_{2n})}$  
measures the interaction strength. Note that
$0<K_{n}\leq 1$ for repulsive interactions;  $K_{n}=1$ 
corresponds to noninteracting systems while
small values of $K_{n}$ indicate a strongly correlated regime.
In our case  $K_{0}=1$ but 
no complications arise from arbitrary values of 
$K_{0}$, which is therefore left unspecified. As we shall see, such freedom permits to 
address general initial-state dependences.

\subsection{Excitation energy}

The excitation energy $\Delta E(t)$ is defined as the difference between 
the energy of the LL at time $t$ and the ground-state energy of the 
LL Hamiltonian at the same time. Since we are interested in $\Delta E(t)$ 
after the interaction quench is completed we consider $t>t_{N-1}$ and 
write
\be
\Delta E (t) = \langle 
\Psi_{0}(t)|  H_{N}| \Psi_{0}(t) \rangle- \langle \Psi_{N}| 
H_{N} | 
\Psi_{N}\rangle ,
\label{exene}
\ee
where $|\Psi_{0} (t) \rangle =  e^{-i 
H_{N}(t-t_{N-1})} \ldots e^{-i 
H_{1}(t_{1}-t_{0})} |  \Psi_{0}  \rangle$ and here and in the 
following $|\Psi_{n}  \rangle$ 
is the ground-state of $H_{n}$.
To calculate $\Delta E (t)$ we  expand the scalar fields in $H_{n}$ as
$
\phi_{\a}(x)=\sum_{q>0}
\frac{i e^{- \frac{aq}{2}} }{\sqrt{2\mathcal{L}q}}[
b^{\dagger}_{\alpha q }e^{-i\e_{\a} qx}-b_{\alpha q}e^{ i\e_{\a}qx}  ] 
+
\frac{\sqrt{\pi}x}{\mathcal{L}}\mathcal{N}_{\alpha},
$
where $\mathcal{N}_{\alpha}$ is the number of
electrons with chirality $\alpha$. Then, the Hamiltonian takes the 
diagonal form 
\be
H_{n}=\sum_{q>0}\sum_{\a}v_{n}q\,b^{\dag}_{\alpha q n}
b_{\alpha q n}+E_{n},
\label{diagham}
\ee
with $E_{n}$  the zero point energy and $b^{(\dagger)}_{\alpha q n}$
the annihilation (creation) operators for elementary excitations of chirality 
$\a$ and momentum $q$  of $H_{n}$. These operators are 
related to the 
non-interacting $b^{(\dagger)}_{\alpha q}$ of the $\f_{\a}$ expansion via the 
Bolgoliubov transformations
$b_{\alpha q n}=b_{\alpha q }
\cosh \varphi_{n} +b^{\dagger}_{\bar{\alpha} q }\sinh \varphi_{n}$, 
with $\varphi_{n} =\frac{1}{2}\tanh^{-1} [(1-K_{n}^{2})/(1+K_{n}^{2}) ]$.
Following the bosonization the average over the ground state $|\Psi_{n} \rangle$ 
is converted into an average over the vacuum
of the $b^{\dagger}_{\alpha q n}$-excitations and hence  $\Delta E (t)$ takes the form
\be
\Delta E (t)= \sum_{q>0} v q \, n_{q}(t),
\label{deltae}
\ee
where  $n_{q}(t)$ is the average of the
number operator $\sum_{\a}b^{\dag}_{\a q N-1}b_{\a q N-1}$ over 
$|\Psi_{0}(t)\rangle$.  
The relation between 
two consecutive boson operators are easily found and read
\be
b_{\alpha q (n+1)}=b_{\alpha q 
n}\cosh \Delta \varphi_{n} +b^{\dagger}_{\bar{\alpha} q 
n}\sinh \Delta \varphi_{n}, 
\label{bog}
\ee
where $\Delta \varphi_{n}=\varphi_{n+1}-\varphi_{n}$.
After a cascade of transformations (\ref{bog}) to express the $b_{\a q 
n}$ in terms of $b_{\a q 0}$ we obtain
$n_{q}(t)\equiv \rho_{q,0}$ as the solution of the 
recursive system of equations 
\beq
P_{q,n}&=&(c_{n}^{2}+s_{n}^{2})P_{q,n+1}+2c_{n}s_{n}\mathrm{Re}[Q_{q,n+1}e^{-i\e_{q}\Delta t_{n} }],
\nonumber \\
Q_{q,n}&=&2c_{n}s_{n}P_{q,n+1}
\nonumber \\
&+&c_{n}^{2}Q_{q,n+1}e^{-i\e_{q}\Delta 
t_{n}}+s_{n}^{2}Q^{*}_{q,n+1}e^{i\e_{q}\Delta t_{n}},
\nonumber \\
\rho_{q,n}&=&\rho_{q,n+1}+2s_{n}^{2}P_{q,n+1}+2 c_{n}s_{n} 
\mathrm{Re}[Q_{q,n+1}e^{-i\e_{q}\Delta t_{n} }],
\nonumber \\
\label{recrelexe}
\eeq
with $\Delta t_{n}=t_{n+1}-t_{n}$, $t_{N}\equiv t$,
$c_{n}=\cosh \Delta\varphi_{n}$, $s_{n}=\sinh \Delta \varphi_{n}$, 
$\e_{q}=2vq$ and boundary conditions $P_{q,N}=1$, 
$Q_{q,N}=\rho_{q,N}=0$. 
In Eq. (\ref{recrelexe}) we assumed that 
$v_{n}=v$ for all $n$; the general 
recursive scheme is simply obtained by replacing $v t_{n}\to 
v_{n}t_{n}$.

\begin{figure}[tbp]
\includegraphics[height=7.cm, width=8cm]{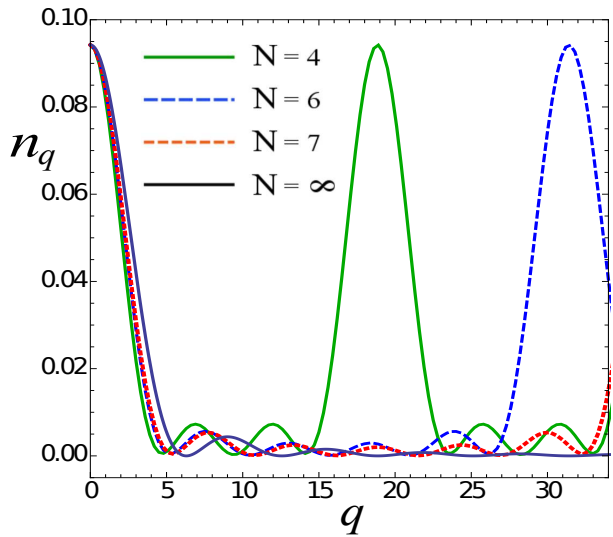}
\caption{Excitation number $n_{q}$ versus $q$ for $N=4,\,6,\,7$ and 
$\inf$ quenches at times $t_{n}=nT/(N-1)$ with 
$K_{n}=1-0.35n/N$ and $T=1$. Momentum $q$ is in units of $1/a$ 
and time in units of $a/v$.}
\label{fig3}
\end{figure}

In Fig. \ref{fig3} we plot $n_{q}(t)$ as a function of $q$ (for 
$t>t_{N-1}=T$ 
there is no dependence on $t$ since the switching protocol finishes 
at $T$) 
for a series of  $N=4,\,6,\,7$ and 
$\inf$ quenches at times $t_{n}=nT/(N-1)$ with $K_{n}=1-0.35n/N$ and $T=1$. 
The distribution of $q$-excitations is peaked at $q=0$ but for any 
finite $N$ there is a revival  every time $q/[2\pi (N-1)]$ is an integer. 
In the next Section we derive an analytic result for $N\to\inf$ 
and show that $n_{q}$ vanishes exponentially at large 
$q$. As $n_{q}$ is not identically zero the LL does not 
thermalize. Below we show how the thermalization breakdown is 
reflected on the equal-time one-particle propagator.

\subsection{The equal-time propagator}

The equal-time one-particle propagator is defined as
\be
G^{[N]}(x,t)=\langle \Psi_{0} (t) | 
\psi_{R}(x)  \psi^{\dag}_{R}(0) |  \Psi_{0} (t) \rangle ,
\label{corr}
\ee
where the superscript $N$ specifies the number of sudden quenches of 
the switching protocol. To calculate $G^{[N]}$ we express the 
fermion fields in terms of the boson fields, expand the latter in 
elementary excitations  and 
exploit the transformations (\ref{bog}) between two consecutive 
$b$ operators. We then obtain
\be
G^{[N]}(x,t) = \frac{1}{2\pi a}
e^{ \sum_{q>0}e^{- aq}\frac{\pi}{\mathcal{L}q} \left( 2i\sin qx  - 
|F_{Rq,0}|^{2}-|F_{Lq,0}|^{2}\right)   } ,
\label{etp}
\ee 
where the functions  $F_{\alpha q,0}(x,t)$ are solutions of the recursive relations
\be
F_{\alpha q,n}=F_{\alpha q,n+1} 
e^{-ivq \Delta t_{n}} c_{n}
-F_{\bar{\alpha}q,n+1} ^{\ast}
e^{ivq\Delta t_{n}} s_{n}, 
\label{recurrel}
\ee
with boundary conditions
$F_{Rq,N}=(e^{iqx}-1) \cosh \varphi_{N}$, 
$F_{Lq,N}=(e^{-iqx}-1)\sinh \varphi_{N}$. Like in Eq. 
(\ref{recrelexe}) the general recursive scheme is obtain by replacing 
$vt_{n}\to v_{n}t_{n}$. 

\begin{figure}[tbp]
\includegraphics[height=7.cm, width=8cm]{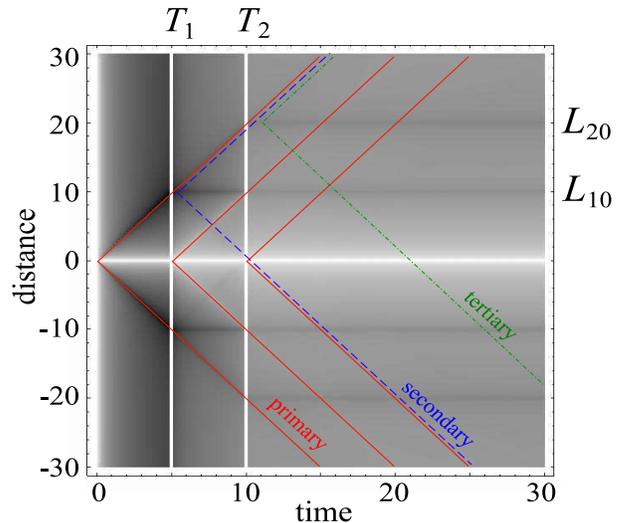}
\caption{Density plot of $\ln |G^{[N]}(x,t)|$
(arbitrary units) for $N=3$ as a function of time and 
distance. The parameters are 
$K_{0}=1,\,K_{1}=0.1,\,K_{2}=0.5,\,K_{3}=0.9$ and 
$T_{0}=0,\,T_{1}=5,\,T_{2}=10$. Time is in units of $10 a/v$ and distance
is in units of $10a$.}
\label{fig1}
\end{figure} 

The full analytic expression of $G^{[N]}$ grows in complexity with 
increasing $N$. To illustrate the typical features of $G^{[N]}$ in 
Fig. \ref{fig1} we report the contour plot of $\ln|G^{[N]}(x,t)|$ for 
$N=3$. The propagator has a Fermi liquid exponent in the light-cone 
region $t < x/2v $, in agreement with Refs. \onlinecite{cazalilla,cardy}.
At the quenching times $t_{n}$ the propagator exhibits a 
time-derivative discontinuity 
due to the sudden change of the interaction which, being a global 
perturbation, instantaneously affects the
whole system. The $n$-th quench gives rise to incoherent excitations 
that propagate at a speed $v$ and those at distance $x$ scatter 
at time $t=t_{n}+x/2v$. When $x=L_{mn}=2v(t_{m}-t_{n})$ the 
scattering time coincides with the time of the $m$-th quench and 
a pronounced peak in $\ln|G^{[N]}(x,t)|$ as a 
function of $x$ develops and  persists forever. The peaks are visible as horizontal lines in the 
contour plot at $x=L_{10}=L_{21}$ and $L_{20}$.
Besides the primary light-cone patterns with origin in $x=0$ and 
$t=t_{n}$ we can see secondary light-cone patterns with origin in 
$x=L_{mn}$ and $t=t_{m}$ which in turn generate ternary light-cone 
patterns and so on and so forth in a cascade that grows like $2^{N}$. 

\begin{figure}[tbp]
\includegraphics[height=7.cm, width=8cm]{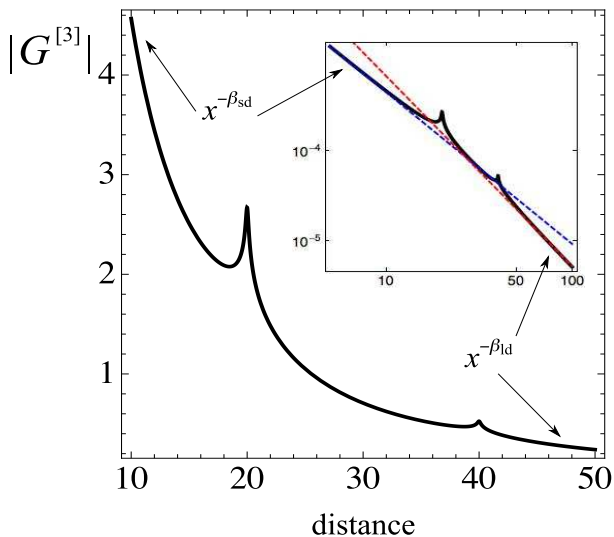}
\caption{Steady-state value of $|G^{[3]}|$ versus $x$ for 
$K_{0}=1,\,K_{1}=0.8,\,K_{2}=0.6,\,K_{3}=0.4$ and 
$T_{0}=0,\,T_{1}=10,\,T_{2}=20$. The crossover between the short and long 
distance regimes is clearly 
visible in the inset.
Distance is in units of $10a$ and
$G^{[3]}$ is in units of $10^{-4}/2\pi a$.}
\label{fig2}
\end{figure}

In Fig. \ref{fig2} we display the steady-state value 
$\lim_{t\to\inf}G^{[3]}(x,t)$  with spikes at the characteristic 
length scales 
$L_{mn}$, in accordance with our previous discussion. 
The analytic expression of $\lim_{t\to\inf}G^{[N]}(x,t)$ is not a simple 
power-law. Nevertheless, a power-law behavior $\sim x^{-\b}$ is 
recovered at short and long distances (see inset of Fig. 
\ref{fig2}). By employing the recursive method of Eq. (\ref{recurrel}) we found 
for the steady-state behavior of $G^{[N]}$ at short distances
\beq
\lim_{x\to 0}\lim_{t\to\inf}G^{[N]}( x,t ) \approx 
\frac{i}{2\pi(x+ia)} \left| \frac{a}{x} 
\right| ^{\beta_{\rm sd}[K_{n}]},
\eeq
with a history and initial-state dependent (through $K_{0}$) exponent
\beq
\beta_{\rm sd}[K_{n}] =  \frac{1+K_{N}^{2}}{2K_{0}}\prod_{n=1}^{N}
\frac{1}{2}\left[1+\left(\frac{K_{n-1}}{K_{n}} \right)^{2} 
   \right] -1.
\label{seq}
\eeq
The product structure of this result is similar to that of the 
exponent $\b$ in the I-V characteristic $I\propto V^{\b}$ of 
an out-of-equilibrium LL subject to a sequence of 
interaction quenches.\cite{perfetto}
Note that $\beta_{\rm sd}$ only depends on the
interaction parameters $K_{n}$ and not on the switching times 
$t_{n}$.\cite{nota2} 
Equation (\ref{seq}) returns the well-known exponent of a LL in the 
ground state for $K_{0}=K_{1}\ldots=K_{N}\equiv K$ since
$\beta_{\rm sd} = (K+1/K)/2-1$, and the exponent of a LL after a single 
interaction quench for $K_{0}=K_{1}=\ldots=K_{n}=1$ and 
$K_{n+1}=\ldots=K_{N}=K$ since $\beta_{\rm sd}=(1 + K^2)^2/(4 K^2)-1$, in 
agreement with Ref. \onlinecite{cazalilla}.

The situation is radically different at long distances. 
In this limit the dependence on $x$ of the steady-state 
propagator is again a power-law
\be
\lim_{x\to \infty}\lim_{t\to\inf}G^{[N]}(x ,t ) \approx 
\frac{i}{2\pi(x+ia)} \left| \frac{R[t_{n},K_{n}]}{x} 
\right| ^{\beta_{\rm ld}[K_{n}]},
\label{ldist}
\ee
but $R[t_{n},K_{n}]$ is a length depending on all intermediate switching
times $t_{n}$ and interactions $K_{n}$. As for the exponent $\b_{\rm ld}$ 
a striking cancellation of the intermediate $K_{n}$ occurs and we find
\be
\beta_{\rm ld}[K_{n}]=\frac{(K_{0}^2 + K_{N}^2)(1 + K_{N}^2) }{4 K_{0} 
K_{N}^2}-1,
\label{long}
\ee
which depends only on the initial state. 

To summarize the steady-state propagator does not thermalize 
for discontinuous switchings, and the thermalization breakdown 
manifests in different ways at short and long distances.
In the next Section we address the evolution of this behavior when 
the sequential protocol approaches a continuous protocol.

\section{Continuous quenches}
\label{contprot}

Continuous switching protocols are obtained as a limiting case of 
a sequential quench. 
Let us start by analyzing again the excitation energy 
$\Delta E(t)$. Taking the quenching times 
$t_{n}=nt/(N-1)$ equally spaced and letting $N\to\inf$ 
the variable $t_{n}$ becomes a continuous variable between $0$ and $t$
and we can construct the differentiable functions $P_{q}(s)$, $Q_{q}(s)$ and 
$\varphi(s)$ according to  
$P_{q}(t_{n})=P_{q,n}$, $Q_{q}(t_{n})=Q_{q,n}$ and 
$\vf(t_{n})=\vf_{n}$. Expanding Eqs. 
(\ref{recrelexe}) to first order in $\D t_{n} = \D t = t/(N-1)$ and $\D 
\vf_{n}$ we  find a coupled system of differential equations
\beq
\frac{d P_{q}}{ds}&=&-2\mathrm{Re}[Q_{q}]\frac{d\vf}{ds},
\nonumber \\
\frac{d Q_{q}}{ds}&=&-2P_{q}\frac{d\vf}{ds} + 2iv q Q_{q},
\label{contexe}
\eeq
that should be  solved with boundary conditions $P_{q}(t)=1$ and $Q_{q}(t)=0$. The 
average occupation number $n_{q}(t)$ is then simply given by $1-P_{q}(0)$.
The most popular continuous protocol is the ramp-switching\cite{tanteramps} 
$\varphi(s)=\bar{\varphi}s/t$ with $0<s<t$. In this case the system (\ref{contexe}) can be solved 
exactly and we find  
\be
n_{q}(t)=\frac{\bar{\varphi}t}{v}\times
\frac{1-\cos
[\frac{2v}{t}\sqrt{q^{2}-(\frac{\bar{\vf}t}{v})^{2}}]}{q^{2}-(\frac{\bar{\vf}t}{v})^{2}},
\label{nqramp}
\ee
which correctly approaches zero for $t\to\inf$ (adiabatic limit), 
whereas for any finite $t$ is exponentially suppressed at large $q$, 
see the curve $N\to\inf$ in Fig. \ref{fig3}. We also 
checked that inserting Eq. (\ref{nqramp}) into Eq. (\ref{deltae}) and 
taking the long-time limit $t\to \inf$ and the weak-coupling limit 
$\bar{\vf}\ll va/t$ the excitation energy vanishes as
\be
\lim_{t\to\inf}\Delta E(t) \sim  \mathcal{L} 
\frac{\ln t }{ t^{2}} ,\quad\quad \bar{\varphi}\ll va/t ,
\label{enelog}
\ee
in agreement with Ref. \onlinecite{dhz.2010}.  
It is interesting to observe that the dependence of $\Delta E\sim 
\d^{2}\ln \d$ on the ramp rate $\d=\bar{\vf}/t$ is not a 
simple power-law $\d^{\nu}$ and, therefore, does 
not belong to the non-analytic regimes contemplated in Ref. \onlinecite{ramp}. 

The system of differential equations (\ref{contexe}) is  exact 
(non-perturbative) and we now exploit it to prove that there exist no 
switching protocol of finite duration capable to drive the initial 
non-interacting system in the interacting ground state of a LL.
Since $\D E(t)$ is the sum of non-negative $n_{q}$'s it is sufficient 
to show that $n_{0}(t)=1-P_{0}(0)$ cannot be zero. The optimal switching
protocol $\vf(s)$ that reproduces a target $P_0(s)$ follows 
from Eqs. (15) with $q=0$; in this case $\vf(s)$ depends
only on the instantaneous value of $P_0(s)$, and reads
\be
\vf(s)=\bar{\vf}-\frac{1}{2}\cosh^{-1} P_{0}(s),
\ee
where $\bar{\vf}$ is the value of the correlation angle at the end of 
the quench.
Thus, $P_{0}(0)=1$ only provided that the initial and final strength of 
the interaction is the same, $\vf(0)=\bar{\vf}$. In particular 
$n_{0}(t)> 0$ for all switching protocols that connect an initial 
interaction strength $K_{0}$ to a final interaction strength $K\neq 
K_{0}$.

Analytic results can be obtained for the one-particle propagator as 
well. Let us start by analyzing the exponents $\b$ of 
the long- and short-distance power-law behavior of $G^{[N]}$. 
For a continuous switching we can construct the
differentiable function $K(s)$ according to 
$K(t_{n})=K_{n}$, with  $K(0)=K_{0}$ and $K(t)=K$.  Approximating 
$K(t_{n-1})=K(t_{n}-\frac{T}{N})\approx 
K(t_{n})-\frac{T}{N}K'(t_{n})$ and taking the logarithm of Eq. 
(\ref{seq}) we find
\begin{eqnarray}
\ln \left[\frac{2K_{0}(\beta_{\rm sd}+1)}{1+K^{2}}\right]&=&
\lim_{N\to \infty}\sum_{n=1}^{N}\log\left[ 1-\frac{T}{N}\frac{K'(t_{n})}{ 
K(t_{n})}\right]
\nonumber \\
&=&-\int_{0}^{t}ds\frac{K'(s)}{K(s)}=\ln\frac{K_{0}}{K},
\nonumber
\end{eqnarray}
and hence
\be
\beta_{\rm sd}[K(s)]=\frac{1}{2}\left(K+\frac{1}{K}\right)-1,
\label{betasd}
\ee
which coincides with the exponent of a ground-state LL. Therefore, 
at short distances the initial-state dependence as well as the history dependence are 
washed out by differentiable switchings and the propagator thermalizes. 
Discontinuous switchings\cite{nota} do, instead, introduce a 
history dependence through the ratio between the values of $K(s)$ 
across the discontinuity. To the contrary the long-distance exponent 
$\b_{\rm ld}$ in Eq. (\ref{long}) is always history independent and  the 
steady-state propagator does never thermalize. Our conclusions agree 
with recent perturbative results by D\`ora et al. in Ref. \onlinecite{dhz.2010}.

\begin{figure}[tbp]
\includegraphics[height=7.cm, width=8cm]{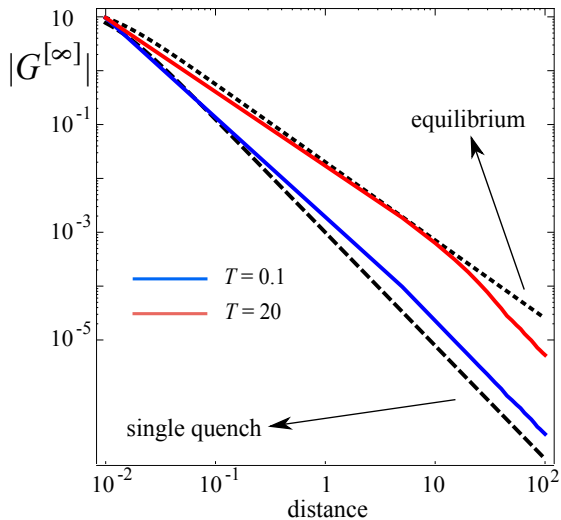}
\caption{Steady-state propagator 
$|G^{[\infty]}|$ versus $x$ for two different ramp
switching-times (in units of $10^{2}a/v$) $T=0.1$ (blue solid) and $T=20$ (red 
solid) and $\bar{\varphi}=0.46$. For comparison the equilibrium 
propagator with exponent $\beta_{\rm{sd}}$ (dotted) and the sudden 
quench propagator with exponent $\beta_{\rm{ld}}$ (dashed) are also 
shown. Distances are in units 
of $10^{2}a$ and $|G^{[\infty]}|$ is 
in units of $10^{-2}/a$.}
\label{fig4}
\end{figure}

To study the crossover from short to long distances we must 
calculate the steady-state propagator at all $x$.
For continuous switching the recursive system of Eq. (\ref{recurrel})
reduces to a system of differential equations 
\be
\frac{d}{ds} F_{\a q}=ivq F_{\a q}+F^{*}_{\bar{\a}q}\frac{d\vf}{ds},
\label{sydeffs}
\ee
that should be solved with boundary conditions 
$F_{Rq}(t)=(e^{iqx}-1)\cosh\vf(t)$ and 
$F_{Lq}(t)=(e^{-iqx}-1)\sinh\vf(t)$. The equal-time propagator 
$G^{[\inf]}(x,t)$ can 
then be calculated from Eq. (\ref{etp}) with $F_{\a q,0}=F_{\a q}(0)$.
In the special case of a ramp protocol $\varphi(s)=\bar{\varphi}s/T$ for $s<T$ and 
$\varphi(s)=\bar{\varphi}$ for $T<s<t$ 
the system of equations (\ref{sydeffs}) can be  solved  analytically and we get
\beq
F_{R q}(0)\!\!&=&\!\!\frac{(1-e^{iqx}) \sinh \gamma_{q} }  {\gamma_{q}} [  
 e^{ivq(t-T)} \bar{\varphi} \sinh 
\bar{\varphi}  \nonumber\\
\!\!&+&\!\! e^{-ivq(t-T)}(   
iqvT -\gamma_{q}\frac{e^{2\gamma_{q}}+1}{e^{2\gamma_{q}}-1}  ) \cosh \bar{\varphi}  ],
\label{fq}
\eeq
where $\gamma_{q}=\sqrt{\bar{\varphi}^{2}-(vqT)^{2}}$; the 
function $F^{*}_{L q}(0)$ is obtained from $F_{R q}(0)$ simply by 
exchanging $\sinh \bar{\varphi} \leftrightarrow \cosh \bar{\varphi}$
and by replacing $v \to -v$.
In Fig. \ref{fig4} we plot the steady-state propagator 
$|G^{[\infty]}|$ as a function of the distance $x$ for two different 
ramp switching-times and $\bar{\varphi}=0.46$, which corresponds to 
the LL parameter $K=0.4$. The results clearly agree with the scenario outlined 
above: at distances smaller than 
the characteristic length $2vT$ the steady-state propagator
behaves thermally while at large distances 
it behaves like the sudden-quench (nonthermal) propagator of Ref. 
\onlinecite{cazalilla}, albeit shifted upwards by a history-dependent 
constant (see also Ref. \onlinecite{dhz.2010}). 

The long-to-short distance crossover is particularly transparent in 
the weak coupling limit $\bar{\varphi}\ll 1$.  In this case we can expand $F_{\a 
q}(0)$ to second order in $\bar{\vf}$, perform the sum over $q$ and 
find the steady-state propagator $G^{[\inf]}$ at all $x$ 
\beq
\lim_{t \to \infty} G^{[\infty]}(x,t ) &\approx& \frac{i}{2\pi (x +ia)} \left | 
\frac{a^{2}}{x^{2}-(2vT)^{2}}\right|^{\bar{\varphi}^{2}}
\left | 
\frac{2vT}{x}\right|^{2\bar{\varphi}^{2}} \nonumber \\
&\times& \left | 
\frac{x^{2}}{x^{2}-(2vT)^{2}}\right|^{\frac{x^{2}\bar{\varphi}^{2}}{4v^{2}T^{2}}}
\left | 
\frac{x-2vT}{x+2vT}\right|^{\frac{x\bar{\varphi}^{2}}{v}}\!\!\!\!,
\label{weak}
\eeq
with power-law exponents depending on $x$. Equation (\ref{weak}) 
reproduces with remarkable accuracy the long-to-short distance 
crossover, which in the same approximation read
\beq
\lim_{x\to 0}\lim_{t \to \infty} G^{[\infty]}(x,t)&\approx& \frac{i}{2\pi (x 
+ia)} \left|\frac{a}{x}\right|^{2\bar{\varphi}^{2}}, \label{pert1}\\
\lim_{x\to \infty}\lim_{t \to \infty} G^{[\infty]}(x,t) &\approx& \frac{i}{2\pi (x 
+ia)} \left|\frac{R}{x}\right|^{4\bar{\varphi}^{2}} ,
\label{pert2}
\eeq
$R=\sqrt{2vTa}$ being a history-dependent length [cfr. Eq.(\ref{ldist})].
The high accuracy stems from the fact that the correlation angle 
$\varphi$ remains small also in the intermediate coupling regime
characterized by $g_{2}\approx g_{4}\approx v_{F}$ for which $\varphi 
\approx 0.1$.

Finally we observe that from Fig. \ref{fig4} and Eq. (\ref{weak}) the 
adiabatic limit is recovered when the switching time $T \to 
\infty$. As $T$ increases the steady-state propagator equals the  
equilibrium propagator up to larger and larger values of $x$.

\section{Conclusions}
\label{concsec}

We provided a comprehensive analysis of the relaxation dynamics of a LL 
after the quenching of the electron-electron interaction for
different switching protocols.
The bosonization method is combined  with a recursive procedure
to address arbitrary sequences of sudden 
quenches and hence, as a limiting case, continuous protocols. 
The approach allows us to evaluate the excitation energy  $\Delta E$ 
and the equal-time one-particle propagator $G$. We found that for a 
sequence of sudden quenches $\D E$ is always larger than zero.
The thermalization breakdown has a dramatic impact on the propagator 
both at finite times and at the steady-state.
In particular the steady-state $G$ exhibits
a power-law behavior $|x|^{-\b}$
with different exponents 
at long and short distances. Remarkable we are able to express $\b$
as explicit functionals of the switching protocol.
We found that at long distances $\b=\beta_{\rm ld}$ carries informations 
on the initial state but not on the history of the switching protocol, and coincides with
the exponent of a sudden quench.\cite{cazalilla} At short distances 
$\b=\beta_{\rm sd}$ is, instead,
both initial-state and history dependent; it is only for continuous 
protocols that memory is washed out and $\beta_{\rm sd}$ equals the 
exponent of a LL in equilibrium.

The continuous limit of the recursive procedure leads to a simple 
system of differential equations  that can be solved numerically. 
For the excitation energy $\Delta E$ we proved that there exist no  
switching protocols of finite duration capable to bring a 
system from a non-interacting ground-state to the ground state of a LL.
It is only in the adiabatic limit that thermalization occurs.
These results may be relevant to design the optimal switching protocol that 
minimizes the total excitation energy or the energy a of 
given $q$-excitation.
For the propagator $G$ we clarify how the adiabatic limit is attained by 
studying ramp-switching protocols of increasing duration $T$.
At weak coupling we derive an explicit expression for $G(x,t\to \infty)$ 
that is valid for all $x$ and around the crossover distance
$x\approx 2vT$ is dominated by a power-law $|x-2vT|^{-\b(x)}$ with a 
$x$-dependent exponent. The formula further reproduces with remarkable 
accuracy the different power-laws at long and short distances.
Our predictions could be experimentally confirmed by measurements in 
ultracold fermionic atoms loaded in optical lattices,
recently proposed\cite{llopt1,llopt2} as candidates to realize 
highly controllable and tunable LLs.



\begin{thebibliography}{99}
\bibitem{nota2}
It is possible to show that the critical exponents $\b_{\rm sd}$ and $\b_{\rm ld}$ 
are independent of the value of the renormalized velocities $v_{n}$.
\bibitem{nota}
For memory effects to be detectable it is enough that $K(t)$ changes on a time 
scale faster than $a/v$.
\end{thebibliography}
\end{document}